\documentclass[11pt]{article}
\usepackage{amsmath,amssymb,amsthm,amsxtra,overpic,bbm,bm,epsfig,subfigure}
\usepackage{color}
\def\thefootnote{\fnsymbol{footnote}}

\begin{document}

\vspace{0.2cm}

\begin{center}
{\Large\bf A partial $\mu$-$\tau$ symmetry and its
prediction for leptonic CP violation}
\end{center}

\vspace{0.2cm}

\begin{center}
{\bf Zhi-zhong Xing $^{a,b}$} \footnote{E-mail: xingzz@ihep.ac.cn}
\quad {\bf Shun Zhou $^{a,c}$} \footnote{E-mail: shunzhou@kth.se;
zhoush@ihep.ac.cn}
\\
{$^a$Institute of High Energy Physics, Chinese Academy of
Sciences, P.O. Box 918 (4), Beijing 100049, China \\
$^b$Center for High Energy Physics, Peking University, Beijing 100080, China, \\
$^c$Department of Theoretical Physics, School of Engineering Sciences,
KTH Royal Institute of Technology, AlbaNova University Center, 10691 Stockholm,
Sweden}
\end{center}

\vspace{1.5cm}
\begin{abstract}
We find that the lepton flavor mixing matrix $U$ should possess a
partial $\mu$-$\tau$ permutation symmetry $|U^{}_{\mu 1}| =
|U^{}_{\tau 1}|$, and the latter predicts a novel correlation
between the Dirac CP-violating phase $\delta$ and three flavor
mixing angles $\theta^{}_{12}$, $\theta^{}_{13}$ and
$\theta^{}_{23}$ in the standard parametrization. Inputting the
best-fit values of these angles reported by Capozzi {\it et al}, we
obtain the prediction $\delta \simeq 255^\circ$ in the normal
neutrino mass ordering, which is in good agreement with the best-fit
result $\delta \simeq 250^\circ$. In this connection the inverted
neutrino mass ordering is slightly disfavored. If this partial
$\mu$-$\tau$ symmetry is specified to be $|U^{}_{\mu 1}| =
|U^{}_{\tau 1}| =1/\sqrt{6}~$, one can reproduce the
phenomenologically-favored relation $\sin^2\theta^{}_{12} = \left(1
- 2\tan^2\theta^{}_{13}\right)/3$ and a viable two-parameter
description of $U$ which were first uncovered in 2006. Moreover, we
point out that the octant of $\theta^{}_{23}$ and the quadrant of
$\delta$ can be resolved thanks to the slight violation of
$|U^{}_{\mu 2}| = |U^{}_{\tau 2}|$ and $|U^{}_{\mu 3}| = |U^{}_{\tau
3}|$ either at the tree level or from radiative corrections.
\end{abstract}

\begin{flushleft}
\hspace{0.8cm} PACS number(s): 14.60.Pq, 13.10.+q, 25.30.Pt
\end{flushleft}

\def\thefootnote{\arabic{footnote}}
\setcounter{footnote}{0}

\newpage

\framebox{\large\bf 1} \hspace{0.1cm} Thanks to a number of successful
experiments which have convincingly
revealed the solar, atmospheric, reactor and accelerator neutrino
(or antineutrino) oscillations \cite{PDG}, the fact that the three
known neutrinos have finite masses and lepton flavors are
significantly mixed has been established on solid ground. While the
origin of neutrino masses requires the existence of new physics
which remains unknown or unsure to us, the leptonic weak
charged-current interactions can be described in the standard way as
follows:
\begin{eqnarray}
-{\cal L}^{}_{\rm cc} = \frac{g}{\sqrt{2}} \ \overline{\left(e
\hspace{0.3cm} \mu \hspace{0.3cm} \tau \right)^{}_{\rm L}} \
\gamma^\mu \left( \begin{matrix} U^{}_{e 1} & U^{}_{e 2} & U^{}_{e
3} \cr U^{}_{\mu 1} & U^{}_{\mu 2} & U^{}_{\mu 3} \cr U^{}_{\tau 1}
& U^{}_{\tau 2} & U^{}_{\tau 3} \cr \end{matrix} \right)
\left(\begin{matrix} \nu^{}_1 \cr \nu^{}_2 \cr \nu^{}_3 \cr
\end{matrix}\right)^{}_{\rm L} W^-_\mu  ~ + ~ {\rm h.c.} \; ,
\end{eqnarray}
where the charged-lepton and neutrino fields are all the mass
eigenstates, and $U$ denotes the $3\times 3$ lepton flavor mixing matrix
\cite{MNS}. If $U$ is unitary, it can be parametrized in terms of
three mixing angles and three CP-violating phases:
\begin{eqnarray}
U = \left( \begin{matrix} c^{}_{12} c^{}_{13} & s^{}_{12} c^{}_{13}
& s^{}_{13} e^{-{\rm i} \delta} \cr -s^{}_{12} c^{}_{23} - c^{}_{12}
s^{}_{13} s^{}_{23} e^{{\rm i} \delta} & c^{}_{12} c^{}_{23} -
s^{}_{12} s^{}_{13} s^{}_{23} e^{{\rm i} \delta} & c^{}_{13}
s^{}_{23} \cr s^{}_{12} s^{}_{23} - c^{}_{12} s^{}_{13} c^{}_{23}
e^{{\rm i} \delta} & -c^{}_{12} s^{}_{23} - s^{}_{12} s^{}_{13}
c^{}_{23} e^{{\rm i} \delta} & c^{}_{13} c^{}_{23} \cr
\end{matrix} \right) P^{}_\nu \; ,
\end{eqnarray}
where $c^{}_{ij} \equiv \cos\theta^{}_{ij}$, $s^{}_{ij} \equiv
\sin\theta^{}_{ij}$ (for $ij = 12, 13, 23$), $\delta$ is often
referred to as the Dirac CP-violating phase, and $P^{}_\nu = {\rm
Diag}\left\{e^{{\rm i}\rho}, e^{{\rm i}\sigma}, 1\right\}$ is a
diagonal Majorana phase matrix which has nothing to do with neutrino
oscillations. So far $\theta^{}_{12}$, $\theta^{}_{13}$ and
$\theta^{}_{23}$ have all been measured to a good degree of
accuracy, and a preliminary hint for a nontrivial value of $\delta$
has also been obtained from a global analysis of current neutrino
oscillation data \cite{FIT}. Here let us quote the results obtained
recently by Capozzi {\it et al} \cite{Fogli} and list them in Table
1, from which the best-fit values of $\theta^{}_{12}$,
$\theta^{}_{13}$, $\theta^{}_{23}$ and $\delta$ are found to be
\begin{eqnarray}
\theta^{}_{12} \simeq 33.7^\circ \; , \hspace{1cm} \theta^{}_{13}
\simeq 8.8^\circ \; , \hspace{1cm} \theta^{}_{23} \simeq 41.4^\circ
\; , \hspace{1cm} \delta \simeq 250^\circ \; ,
\end{eqnarray}
for the normal neutrino mass ordering (i.e., $m^{}_1 < m^{}_2 <
m^{}_3$); or
\begin{eqnarray}
\theta^{}_{12} \simeq 33.7^\circ \; , \hspace{1cm} \theta^{}_{13}
\simeq 8.9^\circ \; , \hspace{1cm} \theta^{}_{23} \simeq 42.4^\circ
\; , \hspace{1cm} \delta \simeq 236^\circ \; ,
\end{eqnarray}
for the inverted neutrino mass ordering (i.e., $m^{}_3 < m^{}_1 <
m^{}_2$). How to understand the observed pattern of lepton flavor
mixing is one of the fundamental issues in particle physics. Before
$\theta^{}_{13}$ was measured, a lot of phenomenological interest
had been paid to the $\mu$-$\tau$ permutation symmetry for model
building. The $\mu$-$\tau$ symmetry is powerful in the sense that it
can constrain the neutrino mass texture and predict $\theta^{}_{23}
= 45^\circ$ together with $\theta^{}_{13} = 0^\circ$ or $\delta =
90^\circ$ (or $270^\circ$), but its validity has been being
challenged since a relatively large value of $\theta^{}_{13}$ was
uncovered by the Daya Bay experiment \cite{DYB} in 2012. Once the
octant of $\theta^{}_{23}$ is fixed and the value of $\delta$ is
measured in the near or foreseeable future, it will be much easier
to probe the possible flavor symmetry behind $U$ and even pin down
the relevant symmetry breaking effects.

In this Letter we stress that even a partial $\mu$-$\tau$ symmetry,
defined as $|U^{}_{\mu i}| = |U^{}_{\tau i}|$ (for $i=1, 2$, or
$3$), is powerful enough for us to establish some testable relations
among the three neutrino mixing angles and the Dirac CP-violating
phase. Note that $|U^{}_{\mu 3}| = |U^{}_{\tau 3}|$ is trivial
because it predicts $\theta^{}_{23} = 45^\circ$, which is apparently
in conflict with the best-fit value of $\theta^{}_{23}$. But we find
that the partial $\mu$-$\tau$ symmetry $|U^{}_{\mu 1}| = |U^{}_{\tau
1}|$ is quite nontrivial and can predict a novel correlation between
$\delta$ and $(\theta^{}_{12}, \theta^{}_{13}, \theta^{}_{23})$,
from which we obtain the prediction $\delta \simeq 255^\circ$ in the
normal neutrino mass ordering. This result is certainly in good
agreement with the best-fit result $\delta \simeq 250^\circ$ given
in Eq. (3). In this connection the inverted neutrino mass ordering
is slightly disfavored. Moreover, $|U^{}_{\mu 2}| = |U^{}_{\tau 2}|$
is phenomenologically disfavored. If the partial $\mu$-$\tau$
symmetry is specified to be $|U^{}_{\mu 1}| = |U^{}_{\tau 1}|
=1/\sqrt{6}$, one can easily reproduce the successful relation
$\sin^2\theta^{}_{12} = \left(1 - 2\tan^2\theta^{}_{13}\right)/3$
and a viable two-parameter description of $U$ which were first
proposed and discussed in 2006. We thus come to an important
conclusion from looking at the global fit of the present neutrino
oscillation data: behind the observed pattern of lepton flavor
mixing should be the partial $\mu$-$\tau$ symmetry $|U^{}_{\mu 1}| =
|U^{}_{\tau 1}|$, and leptonic CP violation must be quite
significant. This observation may serve as one of the guiding
principles for further model building in the lepton sector. We also
point out that the octant of $\theta^{}_{23}$ and the quadrant of
$\delta$ can be resolved thanks to the slight violation of
$|U^{}_{\mu 2}| = |U^{}_{\tau 2}|$ and $|U^{}_{\mu 3}| = |U^{}_{\tau
3}|$ either at the tree level or from radiative corrections.
\begin{table}[t]
\begin{center}
\vspace{-0.25cm} \caption{The best-fit values, together with the
1$\sigma$, 2$\sigma$ and 3$\sigma$ intervals, for the three neutrino
mixing angles and the Dirac CP-violating phase from a global
analysis of current experimental data \cite{Fogli}.} \vspace{0.2cm}
\begin{tabular}{c|c|c|c|c}
\hline
\hline
Parameter & Best fit & 1$\sigma$ range & 2$\sigma$ range & 3$\sigma$ range \\
\hline
\multicolumn{5}{c}{Normal neutrino mass ordering
$(m^{}_1 < m^{}_2 < m^{}_3$)} \\ \hline
$\sin^2\theta_{12}/10^{-1}$
& $3.08$ & 2.91 --- 3.25 & 2.75 --- 3.42 & 2.59 --- 3.59 \\
$\sin^2\theta_{13}/10^{-2}$
& $2.34$ & 2.15 --- 2.54 & 1.95 --- 2.74 & 1.76 --- 2.95 \\
$\sin^2\theta_{23}/10^{-1}$
& $4.37$  & 4.14 --- 4.70 & 3.93 --- 5.52 & 3.74 --- 6.26 \\
$\delta/180^\circ$ &  $1.39$ & 1.12 --- 1.77 & 0.00 --- 0.16
$\oplus$ 0.86 --- 2.00 & 0.00 --- 2.00 \\ \hline
\multicolumn{5}{c}{Inverted neutrino mass ordering
$(m^{}_3 < m^{}_1 < m^{}_2$)} \\ \hline
$\sin^2\theta_{12}/10^{-1}$
& $3.08$ & 2.91 --- 3.25 & 2.75 --- 3.42 & 2.59 --- 3.59 \\
$\sin^2\theta_{13}/10^{-2}$
& $2.40$ & 2.18 --- 2.59 & 1.98 --- 2.79 & 1.78 --- 2.98 \\
$\sin^2\theta_{23}/10^{-1}$
& $4.55$  & 4.24 --- 5.94 & 4.00 --- 6.20 & 3.80 --- 6.41 \\
$\delta/180^\circ$ &  $1.31$ & 0.98 --- 1.60 & 0.00 --- 0.02
$\oplus$ 0.70 --- 2.00 & 0.00 --- 2.00 \\ \hline\hline
\end{tabular}
\end{center}
\end{table}

\vspace{0.5cm}

\framebox{\large\bf 2} \hspace{0.1cm} Without involving any model-dependent uncertainties, let us focus on the $\mu$-$\tau$ permutation symmetry of the
lepton flavor mixing matrix $U$ itself. As already pointed out in
Ref. \cite{XZ2008}, the equalities $|U^{}_{\mu i}| = |U^{}_{\tau i}|$
(for $i=1,2,3$) simultaneously hold if the following conditions are satisfied:
\begin{itemize}
\item     Case A: $\theta^{}_{23} = 45^\circ$ and $\theta^{}_{13} =0^\circ$.
This possibility has been excluded by the Daya Bay experiment
\cite{DYB};

\item     Case B: $\theta^{}_{23} = 45^\circ$ and $\delta =
90^\circ$ or $270^\circ$. This possibility is disfavored but still
allowed by the present experimental data at the $2\sigma$ level, as
one can see from Table 1.
\end{itemize}
Case B is clearly illustrated by the
$\delta$-versus-$\theta^{}_{23}$ plot shown in Fig.~1, in which the
crosspoints $P$ and $Q$ correspond to $\delta =90^\circ$ and
$270^\circ$, respectively. The line dictated by $|U^{}_{\mu 3}| =
|U^{}_{\tau 3}|$ is actually independent of $\delta$ in the given
parametrization of $U$, while the curves dictated by $|U^{}_{\mu 1}|
= |U^{}_{\tau 1}|$ and $|U^{}_{\mu 2}| = |U^{}_{\tau 2}|$ involve
some uncertainties coming from the $1\sigma$ ranges of
$\theta^{}_{12}$ and $\theta^{}_{13}$.
\begin{figure}[t!]
\begin{center}
\subfigure{
\includegraphics[width=0.597\textwidth]{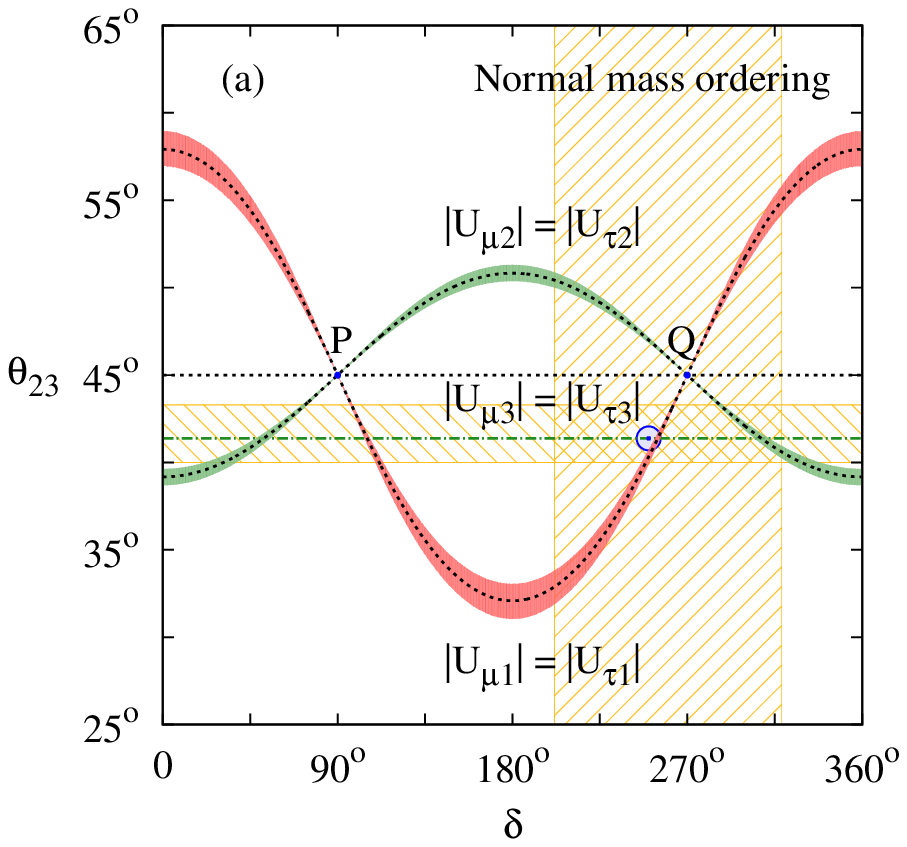} }
\subfigure{
\includegraphics[width=0.597\textwidth]{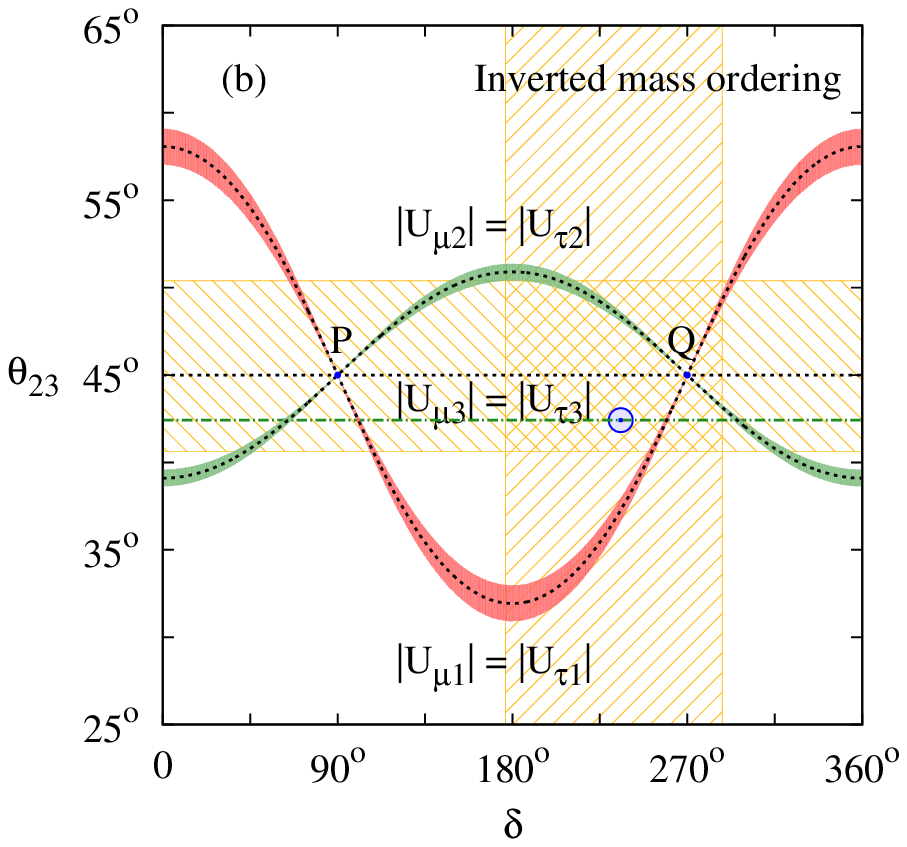} }
\caption{The numerical correlation between $\delta$ and
$\theta^{}_{23}$ as dictated by the partial $\mu$-$\tau$ permutation
symmetry $|U^{}_{\mu i}| = |U^{}_{\tau i}|$ (for $i=1,2,3$) in the
normal or inverted neutrino mass ordering, where the dotted lines
and shaded areas come respectively from inputting the best-fit
values and $1\sigma$ ranges of $\theta^{}_{12}$ and
$\theta^{}_{13}$, while the hashed regions denote the $1\sigma$
ranges of $\delta$ and $\theta^{}_{23}$~\cite{Fogli}. Moreover, the
horizontal dotted-dashed line stands for the best-fit value of
$\theta^{}_{23}$. Three dotted lines cross at points $P$ and $Q$,
corresponding to $(\theta^{}_{23}, \delta) = (45^\circ, 90^\circ)$
and $(45^\circ, 270^\circ)$, respectively. Note that the location of
the best-fit values $(\theta^{}_{23}, \delta) \simeq (41.4^\circ,
250^\circ)$ shown in Fig.~1(a) is marked by the point in the blue
circle and very close to the location of $(\theta^{}_{23}, \delta)
\simeq (41.4^\circ, 255^\circ)$ as predicted by Eq. (5). But in the
inverted neutrino mass ordering case shown in Fig.~1(b), there is an
obvious discrepancy between the best-fit value $\delta \simeq
236^\circ$ and the prediction $\delta \simeq 259^\circ$.}
\end{center}
\end{figure}

When the best-fit values of $\delta$ and $\theta^{}_{23}$ are taken
into account, Fig.~1 shows that the partial $\mu$-$\tau$ symmetry
$|U^{}_{\mu 1}| = |U^{}_{\tau 1}|$ is amazingly favored if the
neutrino mass ordering is normal. This fact is easily understood,
since $|U^{}_{\mu 1}| = |U^{}_{\tau 1}|$ predicts a simple but
nontrivial correlation among $\theta^{}_{12}$, $\theta^{}_{13}$,
$\theta^{}_{23}$ and $\delta$:
\begin{eqnarray}
\cos\delta = \frac{1}{2} \left(\sin^2\theta^{}_{13} -
\tan^2\theta^{}_{12} \right) \cot\theta^{}_{12} \csc\theta^{}_{13}
\cot 2\theta^{}_{23} \; .
\end{eqnarray}
Given $\tan^2\theta^{}_{12} > \sin^2\theta^{}_{13}$, the
CP-violating phase $\delta$ must be in the second or third quadrant
if $\theta^{}_{23} < 45^\circ$ holds. This observation is
qualitatively consistent with the best-fit results of
$\theta^{}_{23}$ and $\delta$ in Eq. (3) or (4). Using the best-fit
values of $\theta^{}_{12}$, $\theta^{}_{13}$ and $\theta^{}_{23}$ as
the inputs, we find that Eq. (5) quantitatively predicts
\begin{eqnarray}
~\delta \simeq \left\{
\begin{array}{lcl}
105^\circ & {\rm or} & 255^\circ \hspace{0.85cm} ({\rm normal ~ mass
~ ordering}) \; , \\ 101^\circ & {\rm or} & 259^\circ \hspace{0.7cm}
({\rm inverted ~ mass ~ ordering}) \; .
\end{array}
\right .
\end{eqnarray}
One can see that the prediction $\delta \simeq 255^\circ$ is in good
agreement with the preliminary best-fit result $\delta \simeq
250^\circ$ in the normal neutrino mass ordering, while the
prediction $\delta \simeq 259^\circ$ in the inverted neutrino mass
ordering is slightly disfavored as compared with the corresponding
best-fit result $\delta \simeq 236^\circ$. Fig.~1 shows the location
of the best-fit points of $\theta^{}_{23}$ and $\delta$. It
coincides with the curve dictated by $|U^{}_{\mu 1}| = |U^{}_{\tau
1}|$ to a good degree of accuracy when the neutrino mass ordering is
normal.

For the sake of comparison, we have also examined the possible
partial $\mu$-$\tau$ symmetries by taking account of a recent and
independent global analysis of the neutrino oscillation data done
by Forero {\it et al} in Ref. \cite{Valle}. The numerical results are shown in Fig.~2. Since the global-fit results obtained in Ref.~\cite{Valle} slightly favor the second octant of $\theta^{}_{23}$, the best-fit value of $\delta$ is closer to the prediction based on the partial $\mu$-$\tau$ symmetry $|U^{}_{\mu 2}| = |U^{}_{\tau 2}|$ in the normal neutrino mass ordering. In the inverted neutrino mass ordering, however,
the best-fit value of $\delta$ is lying between the predictions from
$|U^{}_{\mu 1}| = |U^{}_{\tau 1}|$ and $|U^{}_{\mu 2}| = |U^{}_{\tau 2}|$,
with a slight preference for the former. In this work we concentrate our
attention only on the the partial $\mu$-$\tau$ symmetry
$|U^{}_{\mu 1}| = |U^{}_{\tau 1}|$ and its phenomenological implications.
\begin{figure}[t!]
\begin{center}
\subfigure{
\includegraphics[width=0.597\textwidth]{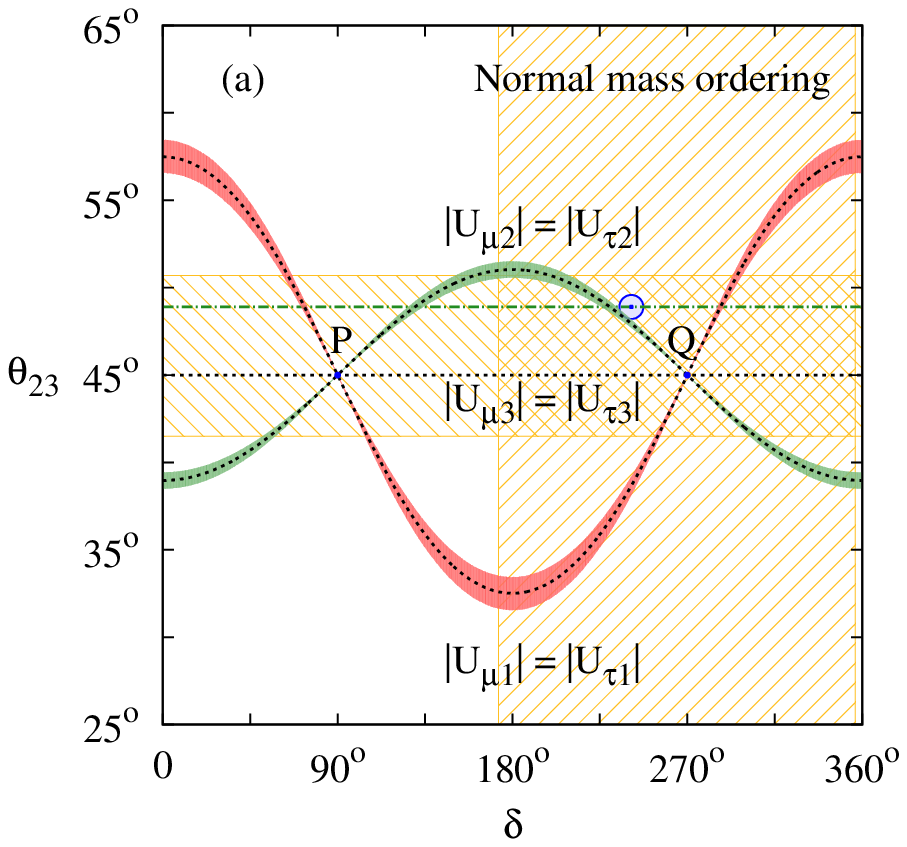} }
\subfigure{
\includegraphics[width=0.597\textwidth]{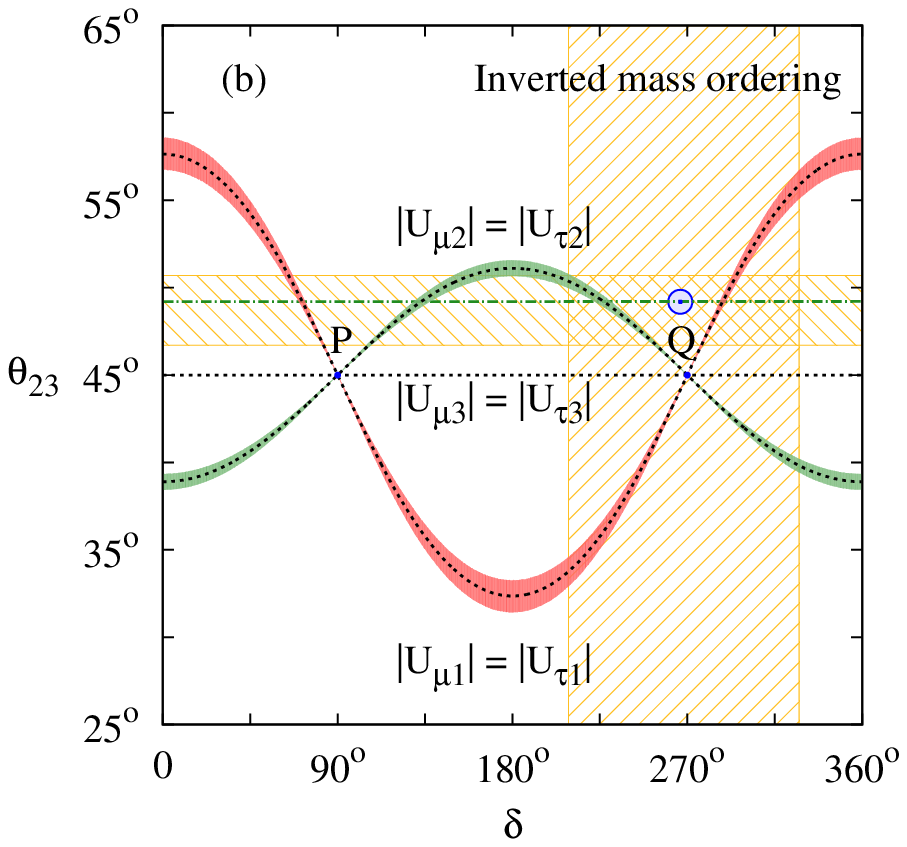} }
\caption{The numerical correlation between $\delta$ and
$\theta^{}_{23}$ as dictated by the partial $\mu$-$\tau$ permutation
symmetry $|U^{}_{\mu i}| = |U^{}_{\tau i}|$ (for $i=1,2,3$) in the
normal or inverted neutrino mass ordering, but for the global-fit data from Ref.~\cite{Valle}. The notations are the same as in Fig.~1. Note that the location of the best-fit values $(\theta^{}_{23}, \delta) \simeq (48.9^\circ, 241^\circ)$ shown in Fig.~2(a) is marked by the point in the blue circle and close to the location of $(\theta^{}_{23}, \delta)
\simeq (48.9^\circ, 231^\circ)$ as predicted by $|U^{}_{\mu 2}| = |U^{}_{\tau 2}|$. But in the inverted neutrino mass ordering case shown in Fig.~2(b), there is an obvious discrepancy between the best-fit value $\delta \simeq 266^\circ$ and the prediction $\delta \simeq 227^\circ$ by $|U^{}_{\mu 2}| = |U^{}_{\tau 2}|$ (or $\delta \simeq 288^\circ$ by $|U^{}_{\mu 1}| = |U^{}_{\tau 1}|$).}
\end{center}
\end{figure}

Note also that the partial $\mu$-$\tau$ permutation symmetry
$|U^{}_{\mu 1}| = |U^{}_{\tau 1}|$ immediately leads us to the relation
\begin{eqnarray}
|U^{}_{\mu 2}|^2 - |U^{}_{\tau 2}|^2 = |U^{}_{\tau 3}|^2 -
|U^{}_{\mu 3}|^2 = \cos^2\theta^{}_{13} \cos 2\theta^{}_{23} =
\cos^2\theta^{}_{13} \sin 2 \epsilon \; ,
\end{eqnarray}
where $\epsilon \equiv 45^\circ - \theta^{}_{23}$ characterizes the
octant of $\theta^{}_{23}$. This result
indicates that the deviation of $\theta^{}_{23}$ from $45^\circ$
actually originates from a slight violation of the partial
$\mu$-$\tau$ symmetry $|U^{}_{\mu 2}| = |U^{}_{\tau 2}|$ or
$|U^{}_{\mu 3}| = |U^{}_{\tau 3}|$. In other words,
$|U^{}_{\tau 3}| > |U^{}_{\mu 3}|$ and
$|U^{}_{\mu 2}| > |U^{}_{\tau 2}|$ hold if $\theta^{}_{23}$ lies in
the first octant (i.e., $\epsilon > 0$); and the
inverse inequalities hold if $\theta^{}_{23}$ lies in
the second octant (i.e., $\epsilon < 0$). Because of
$\cos\delta \propto \tan 2\epsilon$, as one can see from Eq. (5),
the quadrant of $\delta$
is also associated with the sign of $\epsilon$.
In view of the best-fit value of $\theta^{}_{23}$ given in Eq. (3)
or (4), we obtain $\epsilon \simeq 3.6^\circ$ (or $2.6^\circ$)
in the normal (or inverted) neutrino mass ordering. The upcoming
neutrino oscillation experiments will verify or falsify such a
preliminary result and then allow us to resolve the octant of
$\theta^{}_{23}$. On the theoretical side, the smallness of $\epsilon$
can naturally be ascribed to either a tree-level perturbation or
a radiative correction to the exact $\mu$-$\tau$ symmetry.

One might argue that the best-fit result of $\delta$ remains too
preliminary (i.e., $\sin\delta <0$ at the $1.6\sigma$ or $90\%$
confidence level in the normal neutrino mass ordering \cite{Fogli})
to be believable. This is certainly true, but the history of
particle physics tells us that a preliminary result coming from a
reliable global analysis of relevant experimental data often turns
out to be the truth. The closest example of this kind should be the
global-fit ``prediction" for a nonzero and unsuppressed
$\theta^{}_{13}$ \cite{Fogli2008}: $\sin^2\theta^{}_{13} = 0.016 \pm
0.010 ~(1\sigma)$, which proved to be essentially correct after the
first direct measurement of $\theta^{}_{13}$ was reported by the
Daya Bay experiment about three and a half years later. In fact, the
T2K measurement of a relatively strong $\nu^{}_\mu \to \nu^{}_e$
appearance signal \cite{T2K} plays an important role in the global
fit to make $\theta^{}_{13}$ consistent with the Daya Bay result and
drive a slight but intriguing preference for $\delta \simeq 1.5\pi$
\cite{Fogli}. Taking all these things seriously, we expect that the
global fit of current neutrino oscillation data can provide useful
guidance again in connection with the CP-violating phase $\delta$,
especially before a direct measurement of $\delta$ itself is
available.

It is therefore meaningful to stress that the fairly good
consistency of Eq. (5) with the best-fit results of
$\theta^{}_{12}$, $\theta^{}_{13}$, $\theta^{}_{23}$ and $\delta$ in
the normal neutrino mass ordering case might not just be a numerical
accident. Provided Eq. (5) is correct or essentially correct, one
may draw at least two important phenomenological conclusions: 1)
leptonic CP violation in neutrino oscillations must be quite
significant, because its strength, which is measured by the Jarlskog
invariant ${\cal J}^{}_\nu = c^{}_{12} s^{}_{12} c^2_{13} s^{}_{13}
c^{}_{23} s^{}_{23} \sin\delta$ \cite{J}, must be at the percent
level for $\delta \simeq 250^\circ$; 2) behind the observed pattern
of lepton flavor mixing should be the partial $\mu$-$\tau$
permutation symmetry $|U^{}_{\mu 1}| =|U^{}_{\tau 1}|$ or its
approximation. Hence a viable model for neutrino mass generation and
flavor mixing ought to accommodate this partial $\mu$-$\tau$
symmetry at low energies, no matter how complicated its original
flavor symmetry groups are \cite{Review}.

In a specific neutrino mass model, of course, either the full
$\mu$-$\tau$ symmetry or a partial one should impose strong
constraints on the elements $\langle m\rangle^{}_{\alpha\beta}$ (for
$\alpha, \beta = e, \mu, \tau$) of the effective Majorana neutrino
mass matrix $M^{}_\nu = U \widehat{M}^{}_\nu U^T$ with
$\widehat{M}^{}_\nu ={\rm Diag}\{m^{}_1, m^{}_2, m^{}_3\}$. For
instance,
\begin{eqnarray}
\langle m\rangle^{}_{\mu\mu} + \langle m\rangle^{}_{\tau\tau}
- 2\langle m\rangle^{}_{\mu\tau} = \sum^3_{i=1} \left[m^{}_i
\left( U^{}_{\mu i} - U^{}_{\tau i}\right)^2 \right] \;
\end{eqnarray}
holds in general, but the $\mu$-$\tau$ permutation symmetry
$|U^{}_{\mu i}| =|U^{}_{\tau i}|$ may simplify it to some extent,
leading to one or two correlative relations among the three matrix
elements $\langle m\rangle^{}_{\mu\mu}$, $\langle
m\rangle^{}_{\tau\tau}$ and $\langle m\rangle^{}_{\mu\tau}$. The
partial $\mu$-$\tau$ symmetry $|U^{}_{\mu 1}| =|U^{}_{\tau 1}|$
implies that such correlative relations might not be very simple,
but it is still possible to obtain very impressive predictions for
lepton flavor mixing and CP violation. This point has already been
seen in the above discussions, and it will become more transparent
and convincing in a much more specific case to be discussed later
on.

\vspace{0.5cm}

\framebox{\large\bf 3} \hspace{0.1cm} To be more specific, let us focus on the partial $\mu$-$\tau$ symmetry $|U^{}_{\mu 1}| = |U^{}_{\tau 1}| =1/\sqrt{6}$. Then $|U^{}_{e1}| = 2/\sqrt{6}$ can be derived from the unitarity of $U$. Taking $U^{}_{e1} = \cos\alpha$, $U^{}_{\mu 1} = -\sin\alpha \cos\beta$ and $U^{}_{\tau 1} = \sin\alpha \sin\beta$ in a way
analogous to the Kobayashi-Maskawa (KM) parametrization \cite{KM},
we obtain $\alpha = \arctan \left(1/\sqrt{2}\right) \simeq
35.3^\circ$ and $\beta = 45^\circ$. So we are left with a KM-like
parametrization of the lepton flavor mixing matrix $U$, in which another flavor
mixing angle $\theta$ and the Dirac CP-violating phase $\phi$ are
free parameters:
\begin{eqnarray}
U = \left( \begin{matrix} \displaystyle \frac{2}{\sqrt{6}} &
\displaystyle \frac{\cos\theta}{\sqrt{3}} & \displaystyle
\frac{\sin\theta}{\sqrt{3}} e^{-{\rm
i} \phi} \cr \displaystyle -\frac{1}{\sqrt{6}} & \displaystyle
\frac{\cos\theta}{\sqrt{3}} -
\frac{\sin\theta}{\sqrt{2}} e^{{\rm i} \phi} &
\displaystyle \frac{\cos\theta}{\sqrt{2}} + \frac{\sin\theta}{\sqrt{3}} e^{-{\rm
i} \phi} \cr \displaystyle \frac{1}{\sqrt{6}} & \displaystyle
-\frac{\cos\theta}{\sqrt{3}} -
\frac{\sin\theta}{\sqrt{2}} e^{{\rm i} \phi} & \displaystyle
\frac{\cos\theta}{\sqrt{2}} - \frac{\sin\theta}{\sqrt{3}} e^{-{\rm
i} \phi} \cr
\end{matrix} \right) P^{}_\nu \; .
\end{eqnarray}
This interesting neutrino mixing pattern was first proposed by us
\cite{XZ2007} and Lam \cite{Lam} in 2006, and it was later referred
to as the TM1 mixing pattern \cite{TM1}. It can actually be obtained
from the tri-bimaximal neutrino mixing pattern \cite{TB} multiplied
by a complex $(2,3)$ rotation matrix from its right-hand side. Some
of the phenomenological implications of Eq. (8) have already been
explored in the literature (see, e.g., Refs. \cite{XZ2007} and
\cite{Zhou2012}), and it can easily be derived from a given neutrino
mass model by implementing an $A^{}_4$ or $S^{}_4$ flavor symmetry
\cite{Lam,FS}. Comparing this KM-like representation of $U$ with the
standard one in Eq. (2) allows us to reproduce the very striking
prediction \cite{XZ2007}:
\begin{eqnarray}
\sin^2\theta^{}_{12} = \frac{1}{3}\left( 1 - 2\tan^2\theta^{}_{13}
\right) \; .
\end{eqnarray}
Given $\theta^{}_{13} \simeq 8.8^\circ$, for example, Eq. (10) leads
us to $\theta^{}_{12} \simeq 34.3^\circ$, a result which is very
close to the best-fit value of $\theta^{}_{12}$ given in Eq. (3)
--- the deviation is only about $0.6^\circ$! This good agreement is
more intuitive in the $\theta^{}_{12}$-versus-$\theta^{}_{13}$ plot
as shown by Fig.~3, in which the best-fit values and uncertainties
of these two flavor mixing parameters have been taken into account
\footnote{It is worth mentioning that the best-fit result
$\theta^{}_{12} = 34.6^\circ$ obtained from \cite{Valle} coincides
with the prediction of Eq. (10) to a better degree of accuracy.
Nevertheless, the best-fit values of $\theta^{}_{23}$ achieved from
this global analysis and the one done by Gonzalez-Garcia {\it et al}
in Ref. \cite{FIT} prefer the second octant (i.e., $\theta^{}_{23} >
45^\circ$), and hence they are in conflict with the partial
$\mu$-$\tau$ symmetry $|U^{}_{\mu 1}| = |U^{}_{\tau 1}|$ and its
consequences [e.g., Eq. (5)]. While the octant of $\theta^{}_{23}$
remains an open issue today, we believe that it must be associated
with a partial or approximate $\mu$-$\tau$ symmetry. In this sense
we find that $\theta^{}_{23} < 45^\circ$ as obtained by Capozzi {\it
et al} \cite{Fogli} is phenomenologically more favored.}.
Moreover, the deviation of $\theta^{}_{23}$ from $45^\circ$ is
related to $\delta$ in a much simpler way in this specific flavor
mixing scenario:
\begin{eqnarray}
\tan 2\epsilon = - \frac{2 \sin\theta^{}_{13} \sqrt{2 \left( 1 - 3
\sin^2\theta^{}_{13}\right)}}{1 - 5\sin^2\theta^{}_{13}} \cos\delta
\; .
\end{eqnarray}
This relation allows us to obtain $\delta \simeq 254^\circ$ from
$\theta^{}_{13} \simeq 8.8^\circ$ and $\epsilon \simeq 3.6^\circ$ in
the normal neutrino mass ordering, compatible with the best-fit
result $\delta \simeq 250^\circ$ in Eq. (3) to a good degree of
accuracy.
\begin{figure}[t!]
\begin{center}
\includegraphics[width=0.7\textwidth]{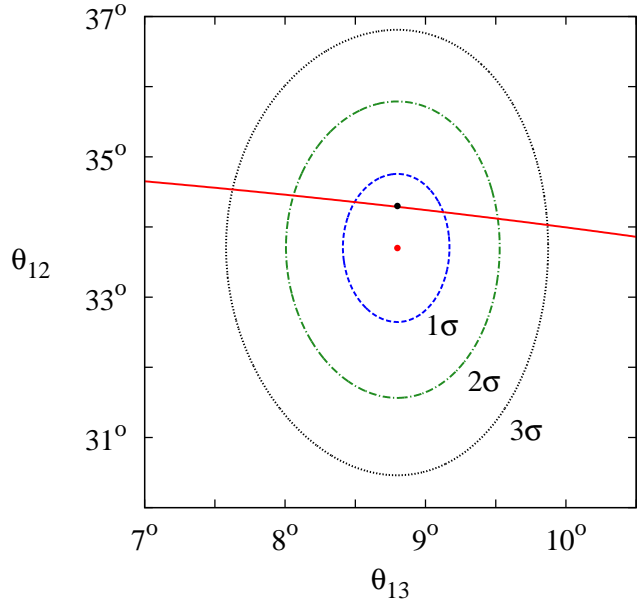}
\vspace{-0.4cm} \caption{A comparison between the relation $\sin^2
\theta^{}_{12} = (1 - 2\tan^2 \theta^{}_{13})/3$ (red line) and the
best-fit values of $\theta^{}_{12}$ and $\theta^{}_{13}$. The
allowed $1\sigma$, $2\sigma$ and $3\sigma$ regions of both
parameters are constructed from their global-fit results in the
normal neutrino mass ordering \cite{Fogli}, and a Gaussian
distribution of the uncertainties is assumed for illustration. If
$\theta^{}_{13} \simeq 8.8^\circ$ is input, one will arrive at the
prediction $\theta^{}_{12} \simeq 34.3^\circ$ (black dot), which is
consistent with the best-fit result $\theta^{}_{12} \simeq
33.7^\circ$ (red dot) within $1\sigma$.}
\end{center}
\end{figure}

We conclude that Eq. (9) should nowadays be the simplest and most
favored neutrino mixing pattern, which respects the partial
$\mu$-$\tau$ permutation symmetry $|U^{}_{\mu 1}| = |U^{}_{\tau 1}|$
and may originate from a violation of the full $\mu$-$\tau$ symmetry
or discrete flavor symmetries. Some comments in this connection are
in order.
\begin{itemize}
\item      {\it Soft breaking of $\mu$-$\tau$ symmetry.} As indicated
by Eq. (9), the lepton flavor mixing matrix $U$ with the partial $\mu$-$\tau$
symmetry $|U^{}_{\mu 1}| = |U^{}_{\tau 1}|$ can be regarded as the
tri-bimaximal mixing pattern corrected by a $(2,3)$ rotation, which
can be induced by a $\mu$-$\tau$ symmetry breaking term in the
neutrino mass matrix $M^{}_\nu$. In the canonical type-I seesaw
model, this soft term may come from the heavy Majorana neutrino mass
matrix; while in the type-II seesaw model, it may arise from the
breaking of $\mu$-$\tau$ symmetry in the potential of triplet and
doublet Higgs fields.

\item      {\it Discrete flavor symmetries.} As a general guiding
principle, one may search for a proper flavor symmetry by
investigating the residual symmetry group $G^{}_\ell$ in the
charged-lepton sector and $G^{}_\nu$ in the neutrino sector
\cite{Lam}. In the basis where the charged-lepton mass matrix is
diagonal, for instance, the residual symmetry in the charged-lepton
sector is $G^{}_\ell = Z^{}_3$, while that in the neutrino sector is
$G^{}_\nu = Z^{}_2$ as implied by the first column of $U$ in Eq.
(9). Therefore, a finite discrete symmetry group which contains
$G^{}_\ell$ and $G^{}_\nu$ as the subgroups can be a good starting
point for model building to achieve the partial $\mu$-$\tau$
permutation symmetry under consideration \cite{FS}.

\item      {\it Radiative breaking of $\mu$-$\tau$ symmetry.}
A specific flavor symmetry model, which can simultaneously predict
$\theta^{}_{23} = 45^\circ$ and $\delta = 90^\circ$ or $270^\circ$
\cite{FS2}, is very likely to work at a superhigh-energy scale
(e.g., close to the grand unification scale $\Lambda \sim 10^{16}$
GeV or around the conventional seesaw scale $\Lambda \sim 10^{14}$
GeV). In this case it is possible to preserve the partial
$\mu$-$\tau$ symmetry $|U^{}_{\mu 1}| = |U^{}_{\tau 1}|$ but break
the equalities $|U^{}_{\mu 2}| = |U^{}_{\tau 2}|$ and $|U^{}_{\mu
3}| = |U^{}_{\tau 3}|$ at the electroweak scale by taking into
account the renormalization-group running effects on the lepton flavor
mixing matrix
$U$. In other words, the octant of $\theta^{}_{23}$ and the quadrant
of $\delta$ can be resolved with the help of radiative corrections
to the full $\mu$-$\tau$ symmetry.
\end{itemize}
So there is a lot of room for model building to understand why the
observed pattern of lepton flavor mixing exhibits an approximate
$\mu$-$\tau$ permutation symmetry.

\vspace{0.5cm}

\framebox{\large\bf 4} \hspace{0.1cm} To summarize, we have taken a close and serious look at the global-fit results of current neutrino oscillation data and found that the partial $\mu$-$\tau$ permutation symmetry $|U^{}_{\mu 1}| = |U^{}_{\tau 1}|$ should be behind the observed pattern of lepton flavor mixing.
This simple but interesting symmetry predicts a novel
correlation between the Dirac CP-violating phase $\delta$ and three
flavor mixing angles $\theta^{}_{12}$, $\theta^{}_{13}$ and
$\theta^{}_{23}$ in the standard parametrization, as shown in Eq.
(5). Inputting the best-fit values of those mixing angles as
reported by Capozzi {\it et al} \cite{Fogli}, we have obtained the
prediction $\delta \simeq 255^\circ$ in the normal neutrino mass
ordering, which is in good agreement with the best-fit result
$\delta \simeq 250^\circ$. In comparison, the inverted neutrino mass
order case is slightly disfavored in this connection. Specifying
this partial $\mu$-$\tau$ symmetry to be $|U^{}_{\mu 1}| =
|U^{}_{\tau 1}| =1/\sqrt{6}$, we have reproduced (Fig.~3) the remarkable
relation $\sin^2\theta^{}_{12} = \left(1 -
2\tan^2\theta^{}_{13}\right)/3$ which was first uncovered in 2006
and is consistent very well with the present experimental data. In
this case we are also left with a viable two-parameter description
of the lepton flavor mixing matrix $U$, and it can easily be derived
from some neutrino mass models with concrete flavor symmetries. Moreover, we
have pointed out that the octant of $\theta^{}_{23}$ and the quadrant of
$\delta$ can be resolved thanks to the slight violation of
$|U^{}_{\mu 2}| = |U^{}_{\tau 2}|$ and $|U^{}_{\mu 3}| = |U^{}_{\tau
3}|$ either at the tree level or from radiative corrections.

Even though the normal neutrino mass ordering is favored in our
discussions, it does not necessarily mean that it would be hopeless
to observe a signal of the neutrinoless double-beta decay. The
reason is simply that a normal but weakly hierarchical (or even
nearly degenerate) neutrino mass spectrum can also result in a
promising value of the effective Majorana neutrino mass term
$|\langle m\rangle^{}_{ee}|$, which is actually comparable with the
expected value in the inverted neutrino mass ordering case
\cite{Rode}.

On the other hand, $\delta \simeq 250^\circ$ implies a very significant
effect of leptonic CP violation, which will be observed in the
next-generation long-baseline neutrino oscillation experiments. One
probably questions the preliminary hint of $\delta \simeq 250^\circ$
as obtained from a global analysis of the available neutrino
oscillation data, but a fairly good lesson learnt recently from the
global-fit determination of $\theta^{}_{13}$ is so encouraging that
the situation for $\delta$ deserves very serious attention. A
similar argument works for the octant of $\theta^{}_{23}$, which
must be associated with a partial or approximate $\mu$-$\tau$
symmetry. That is why we are strongly motivated to highlight the
partial $\mu$-$\tau$ symmetry $|U^{}_{\mu 1}| = |U^{}_{\tau 1}|$ and
its striking prediction for CP violation in the lepton sector. They
will soon be tested by the upcoming experimental data.

\vspace{0.5cm}

We are grateful to T. Ohlsson and other organizers of the Nordita
activity ``News in Neutrino Physics" for warm hospitality in
Stockholm, where this work was initiated. The research of Z.Z.X. was
supported in part by the National Natural Science Foundation of
China under grant No. 11135009.

\end{document}